\documentclass{ifacconf}

\usepackage{graphicx}      
\usepackage{natbib}        
\usepackage[english]{babel}
\begin{document}

\begin{frontmatter}

\title{Fractional relaxation and wave equations for dielectrics characterized by the~Havriliak-Negami response function\thanksref{footnoteinfo}}

\thanks[footnoteinfo]{Author is grateful to the Russian
Foundation for Basic Research (grant~10-01-00608) for financial
support.}

\author[First]{Renat T. Sibatov},
\author[First]{Dmitry V. Uchaikin}

\address[First]{Ulyanovsk State University,
   Russia \\E-mail: ren\_sib@bk.ru, uchaikin@gmail.com}

\begin{abstract}

\vspace{0.2cm}

A fractional relaxation equation in dielectrics with response
function of the Havriliak-Negami type is derived. An explicit
expression for the fractional operator in this equation is
obtained and Monte Carlo algorithm for calculation of action of
this operator and is constructed. Relaxation functions calculated
numerically according to this scheme coincide with analytical
functions obtained  earlier by other authors. The algorithm
represents a numerical way of calculation in relaxation problems
with arbitrary initial and boundary conditions. A fractional
equation for electromagnetic waves in such dielectric media is
obtained. Numerical results are in a good agreement with
experimental data.

\end{abstract}


\end{frontmatter}


\section{Introduction}

When a step-wise electric field is applied, polarization of a
material approaches its equilibrium value not instantly but after
some time later. Hereditary effect of polarization is expressed
through the integral relation. For electric induction in an
isotropic medium, we have [\cite{Fro:58}]
\begin{equation}\label{eq_material_equation}
\mathbf{D}(t)=\varepsilon_\infty\ \mathbf{E}(t)+
\int\limits_{-\infty}^t \kappa(t-t')\ \mathbf{E}(t') d t'.
\end{equation}
Fourier transformation of the last expression gives
$$
\tilde{\mathbf{D}}(\omega)=\varepsilon^*(\omega)\tilde{\mathbf{E}}(\omega),
\quad
\varepsilon^*(\omega)=\varepsilon_\infty+\tilde{\kappa}(\omega)
$$

Relaxation properties of different media (dielectrics,
semiconductors, ferromagnetics, and so on) are normally expressed
in terms of time-domain response function $\phi(t)$ which
represents the current flowing under the action of a step-function
electric field, or of the frequency-dependent real and imaginary
components of its Fourier transform:
$$
\widetilde{\phi}(\omega)=\int\limits_0^\infty e^{-i\omega t}
\phi(t) dt=\phi'(\omega)-i\phi''(\omega).
$$

The Fourier transform of the response function $\phi(t)$ is
related to the frequency dependence of dielectric permittivity
through the following relation
$$
\tilde{\phi}(\omega)=\frac{\varepsilon^*(\omega)-\varepsilon_\infty}{\varepsilon_s-\varepsilon_\infty},\quad
\tilde{\kappa}(\omega)=(\varepsilon_s-\varepsilon_\infty)\tilde{\phi}(\omega).
$$
Here $\varepsilon_s=\varepsilon^*(0)$ is the stationary dielectric
permittivity.

The classical Debye expression (see \cite{Deb:54}) for a system of
noninteracting randomly oriented dipoles freely floating in a
neutral viscous liquid is
$$
\widetilde{\phi}(\omega)=\frac{1}{1+i\omega\tau},
$$
where $\tau$ is the temperature-dependent relaxation time
characterizing the Debye process:
$$
\phi(t)=\phi(0)\exp(-t/\tau),\quad t>0.
$$
The latter function obeys the simple differential equation
$$
\frac{d \phi(t)}{dt}+\tau^{-1} \phi(t)=0.
$$
Numerous experimental data gathered, for instance, in books
(\cite{Jon:77}, \cite{Ram:87}) convincingly show that this theory
is not able to describe relaxation processes in solids, the
relaxation behavior may deviate considerably from the exponential
Debye pattern and exhibits a broad distribution of relaxation
times. There exist a few other empirical response functions for
solids: the Cole–Cole function (\cite{Col:41})
\begin{equation}\label{eq_CC}
\tilde{\phi}(\omega)=\frac{1}{1+(i\omega/\omega_p)^\alpha},\qquad
0<\alpha<1,
\end{equation}
the Cole-Davidson function (\cite{Col:51})
\begin{equation}\label{eq_CD}
\tilde{\phi}(\omega)=\frac{1}{(1+(i\omega/\omega_p))^\beta},\qquad
0<\beta<1,
\end{equation}
and the Havriliak-Negami function (\cite{Hav:66})
\begin{equation}\label{eq_HN}
\tilde{\phi}(\omega)=\frac{1}{(1+(i\omega/\omega_p)^\alpha)^\beta},\qquad
0<\alpha,\beta<1.
\end{equation}
Here $\omega_p$ is the peak of losses, $\alpha$ and $\beta$ are
constant parameters.

In the present paper, we obtain a fractional relaxation equation
in dielectrics with response function of the Havriliak-Negami
type. With the help of definitions of the functional analysis we
derive the explicit expression for fractional operator in this
equation. Then we construct a Monte Carlo algorithm for
calculation of action of this operator and of the inverse one. The
algorithm represents a numerical way of calculation in relaxation
problems with arbitrary initial and boundary conditions. Then, we
consider propagation of electromagnetic waves in such dielectric
media and discuss memory phenomenon.

\section{Havriliak-Negami's relaxation}

The most general approximation for frequency dependence of
response function is given by two-parameter formula proposed by
\cite{Hav:66}. The solution of the corresponding fractional
differential equation
\begin{equation}\label{eq_frac_relax}
\left[1+(\tau\ _0\textsf{D}_t)^\alpha\right]^\beta
\phi(t)=\delta(t),\qquad \tau=1/\omega_p,
\end{equation}
based on the expansion of fractional power of operator sum into
infinite Newton's series
$$
\left[1+(\tau\
_0\textsf{D}_t)^\alpha\right]^\beta=\sum\limits_{n=0}^\infty{\beta\choose
n} \left(\tau\ _0\textsf{D}_t\right)^{\alpha(\beta-n)},
$$
has been obtained by \cite{Nov:05}:
$$
\phi(t)=-\frac{1}{\Gamma(\beta)}\sum\limits_{n=0}^\infty\frac{(-1)^n\Gamma(n+\beta)}
{n!\Gamma(\alpha(n+\beta))}\left(\frac{t}{\tau}\right)^{\alpha(n+\beta)}.
$$

The operator $\left[1+\omega_p^{-\alpha}(
_0\textsf{D}_t)^\alpha\right]^\beta$ can also be presented in the
form (\cite{Nig:97}):
$$
\textsf{W}^{\alpha,\beta}_{\omega_p} f(t)=[1+\omega_p^{-\alpha}\
_{0}\textsf{D}_t^\alpha]^{\beta}f(t)=
$${\small
$$
\omega_p^{-\alpha\beta}\exp\left(-\frac{\omega_p^\alpha
t}{\alpha}\ _0\textsf{D}^{1-\alpha}_t\right)\
_{0}\textsf{D}^{\alpha\beta}_t\exp\left(\frac{\omega_p^\alpha
t}{\alpha}\ _0\textsf{D}^{1-\alpha}_t\right)f(t)
$$}

The HN function is considered as a general expression for the
universal relaxation law (\cite{Jon:96}). This universality is
observed in dielectric relaxation in dipolar and nonpolar
materials, conduction in hopping electronic semiconductors,
conduction in ionic conductors, trapping in semiconductors, decay
of delayed luminescence, surface conduction on insulators,
chemical reaction kinetics, mechanical relaxation, magnetic
relaxation. Despite of quite different intrinsic mechanisms, the
processes manifest astonishing similarity (Jonscher, 1996). The
situation seems to be similar to diffusion processes. Random
movements of small pollen grain visible under a microscope,
neutrons in nuclear reactors, electrons in semiconductors are
quite different processes from physical point of view but they are
the same process of Brownian motion from stochastic point of view.
This analogy stimulates search of an appropriate stochastic model
for the universal relaxation law. Investigations of such kind have
been carried out in the works (\cite{Wer:96}, \cite{Wer:91},
\cite{Nig:84} \cite{Nig:97}, \cite{Glo:93}, \cite{Jur:00},
\cite{Cof:02}, \cite{Dej:03}, \cite{Ayd:05}).

Weron \& Jurlewicz (2005) have shown how to modify the random-walk
scheme underlying the classical Debye response in order to derive
the empirical Havriliak-Negami function. Moreover, they have
derived formulas for simulation of random variables with
probability density function, Fourier transform of which is the
Havriliak-Negami function. These relations contain stable random
variables.

\cite{Cof:02} reformulated the Debye theory of dielectric
relaxation of an assembly of polar molecules using a fractional
noninertial Fokker–Planck equation to explain anomalous dielectric
relaxation.

\cite{Dej:03} considered the fractional approach to the
orientational motion of polar molecules acted on by an external
perturbation. The problem is treated in terms of noninertial
rotational diffusion (configuration space only) which leads to
solving a fractional Smoluchowski equation. This model is in a
good agreement with experimental data for the third-order
nonlinear dielectric relaxation spectra of a ferroelectric liquid
crystal.

Here with the help of relations of the functional analysis for
fractional powers of operators, we derive a new numerical
algorithm of solution of fractional equations corresponding to the
Havriliak-Negami response. This algorithm is based on Monte Carlo
simulation of one-sided stable random variables.

\section{Fractional operator corresponding to the Havriliak-Negami  response}

Let the equi-continuous semigroup $\{T_t; t\geq0\}$ of $C_0$-class
be defined on locally convex linear topological B-space $F$. The
infinitesimal generating operator $A$ of the semigroup $\{T_t\}$
is defined as
$$
A f=\lim\limits_{h\downarrow0}\ h^{-1}(T_h-I)\ f
$$
with domain
$$
D(A)=\{f\in F; \lim\limits_{h\downarrow0}\  h^{-1}(T_h-I)\ f \ \
\mbox{exists in}\ F \}.
$$
According to S. Bochner and R. S. Phillips, the operators
$$
\tilde{T}_{t,\alpha}\ x\equiv\tilde{T}_t \ x=\cases{
\int\limits_0^\infty t^{-1/\alpha}g^{(\alpha)}(t^{-1/\alpha}s)\
T_s x\ ds,& $t>0$ ,\cr \qquad\qquad\quad x,& $t=0$ \cr},
$$
where $g^{(\alpha)}(t)$ is a one-sided stable density, constitute
an equi-continuous semigroup of $C_0$-class.

The corresponding infinitesimal operators are connected through
the following relation (Iosida (1980)):
$$
\tilde{A}_\alpha\ x=-(-A)^\alpha x,\qquad \forall x\in D(A).
$$

The infinitesimal operator $-[1+ _{-\infty}\textsf{D}^\alpha_t]$
generates the semigroup
$$
T_h=e^{-h} e^{-h _{-\infty}\textsf{D}^\alpha_z},
$$
According to the Bochner-Philips relation, the semigroup generated
by the infinitesimal operator $[1+
_{-\infty}\textsf{D}^\alpha_t]^\beta$, where $\beta<1$, has the
form
$$
\widehat{T}_t \ f= \int\limits_0^\infty
t^{-1/\beta}g^{(\alpha)}(t^{-1/\beta}\tau)\ \widetilde{T}_\tau f\
d\tau=
$$
$$
\int\limits_0^\infty d\tau\ e^{-\tau}\
t^{-1/\beta}g^{(\beta)}(t^{-1/\beta}\tau)\int\limits_0^\infty
\tau^{-1/\alpha} g^{(\alpha)}(\tau^{-1/\alpha}u)\ T_u  f\ du
$$
Considering this integral as averaging over ensembles of stable
random variables, we obtain the following relationship
$$
\widehat{T}_t \ f=\left\langle
\exp\left(-t^{1/\beta}S_\beta\right)f\left(z-\left[t^{1/\beta}S_\beta
\right]^{1/\alpha}S_\alpha\right)\right\rangle
$$
From this semigroup we can obtain the corresponding infinitesimal
operator $[1+ _{-\infty}\textsf{D}^\alpha_t]^\beta$.

To find the inverse operator
$$
[1+_{-\infty}\textsf{D}^\alpha_t]^{-\beta},\qquad
0<\alpha,\beta<1,
$$
we use the relation for potential operator
$$
A^{-1} f=\int\limits_0^\infty T_s f ds.
$$
Consequently,
$$
[1+ _{-\infty}\textsf{D}^\alpha_t]^{-\beta}f=
$$
$$
=\left\langle\ \int\limits_0^\infty
\exp\left(-t^{1/\beta}S_\beta\right)\
f\left(z-\left(t^{1/\beta}S_\beta\right)^{1/\alpha}S_\alpha\right)
dt\right\rangle.
$$
Here $S_\alpha$ and $S_\beta$ are one-sided stable random
variables with characteristic exponents $0<\alpha\leq1$ and
$0<\beta\leq1$.

Introducing exponentially distributed random variable $U$, we
arrive at
$$
[1+ _{-\infty}\textsf{D}^\alpha_t]^{-\beta}f=\left\langle\
\int\limits_0^\infty e^{-\xi}\ f\left(z-S_\alpha\xi^{1/\alpha}
\right)
\frac{\beta\xi^{\beta-1}}{S_\beta^\beta}d\xi\right\rangle=
$$
\begin{equation}\label{eq_frac_operator}
=\beta\left\langle S_\beta^{-\beta}U^{\beta-1}
 f\left(z-S_\alpha U^{1/\alpha} \right)
 \right\rangle
\end{equation}

We use this formula to find the solution of fractional relaxation
equation for arbitrary prehistories of charging-discharging
process.

\section{Fractional wave equation}

Substituting the Havrilyak-Negami dependence of permittivity on
frequency
$$
\varepsilon^*(\omega)=\varepsilon_\infty+\frac{\varepsilon_s-\varepsilon_\infty}{[1+(i\omega/\omega_p)^\alpha]^\beta}
$$
into the Fourier transform of the equation
(\ref{eq_material_equation}), we obtain
$$
\mathbf{D}(t)=\varepsilon_\infty\mathbf{E}(t)+(\varepsilon_s-\varepsilon_\infty)\
\omega_p^{\alpha\beta}\times
$$
$$
\exp\left(-\frac{\omega_p^\alpha}{\alpha}\ t\
_{-\infty}\textsf{D}^{1-\alpha}_t\right)\
_{-\infty}\textsf{I}^{\alpha\beta}_t\exp\left(\frac{\omega_p^\alpha}{\alpha}\
t\ _{-\infty}\textsf{D}^{1-\alpha}_t\right)\mathbf{E}(t)
$$
Here special forms of fractional operators arise
$$
\textsf{W}^{\alpha,\beta}_{\omega_p} f(t)=[1+\omega_p^{-\alpha}\
_{-\infty}\textsf{D}_t^\alpha]^{\beta}f(t)=
$${\small
$$
\omega_p^{-\alpha\beta}\exp\left(-\frac{\omega_p^\alpha
t}{\alpha}\ _{-\infty}\textsf{D}^{1-\alpha}_t\right)\
_{-\infty}\textsf{D}^{\alpha\beta}_t\exp\left(\frac{\omega_p^\alpha
t}{\alpha}\ _{-\infty}\textsf{D}^{1-\alpha}_t\right)f(t)
$$}
The inverse operator has the form
$$
\textsf{W}^{\alpha,-\beta}_{\omega_p} f(t)=[1+\omega_p^{-\alpha}\
_{-\infty}\textsf{D}_t^\alpha]^{-\beta}f(t)=
$${\small
$$
=\omega_p^{\alpha\beta}\exp\left(-\frac{\omega_p^\alpha}{\alpha}\
t\ _{-\infty}\textsf{D}^{1-\alpha}_t\right)\
_{-\infty}\textsf{I}^{\alpha\beta}_t\exp\left(\frac{\omega_p^\alpha}{\alpha}\
t\ _{-\infty}\textsf{D}^{1-\alpha}_t\right)f(t)
$$}

The following asymptotical relationships take place:
\begin{equation}\label{eq_asymptotics_HN_operator1}
\textsf{W}^{\alpha,\beta}_{\omega_p}
f(t)\sim\cases{[1+\beta\omega_p^{-\alpha}\
_{-\infty}\textsf{D}^{\alpha}_t] f(t),& $t\gg 1/\omega_p$,\cr &
\cr \omega_p^{-\alpha\beta}\ _{-\infty}\textsf{D}^{\alpha\beta}_t
f(t),& $t\ll 1/\omega_p,$ \cr}
\end{equation}

\begin{equation}\label{eq_asymptotics_HN_operator2}
\textsf{W}^{\alpha,-\beta}_{\omega_p}
f(t)\sim\cases{[1-\beta\omega_p^{-\alpha}\
_{-\infty}\textsf{D}^{\alpha}_t] f(t),& $t\gg 1/\omega_p$,\cr &
\cr \omega_p^{\alpha\beta}\ _{-\infty}\textsf{I}^{\alpha\beta}_t
f(t),& $t\ll 1/\omega_p,$ \cr}
\end{equation}

Maxwell's equations
$$
\mathrm{rot}\
\mathbf{H}=\frac{4\pi}{c}\mathbf{j}+\frac{1}{c}\frac{\partial
\mathbf{D}}{\partial t}
$$
$$
\mathrm{rot}\ \mathbf{E}=-\frac{1}{c}\frac{\partial
\mathbf{B}}{\partial t}
$$
in combination with the material relations
$$
\mathbf{D}=\varepsilon_\infty
\mathbf{E}+(\varepsilon_s-\varepsilon_\infty)[1+\omega_p^{-1}\
_{-\infty}\textsf{D}_t^\alpha]^{-\beta}\ \mathbf{E},\quad
\mathbf{B}=\mu \mathbf{H}
$$
lead to the following wave equation
$$
\frac{\mu\varepsilon_\infty}{c^2}\frac{\partial^2
\mathbf{E}}{\partial
t^2}+\frac{\mu(\varepsilon_s-\varepsilon_\infty)}{c^2}[1+\omega_p^{-1}\
_{-\infty}\textsf{D}_t^\alpha]^{-\beta}\ \frac{\partial^2
\mathbf{E}}{\partial t^2}+
$$
\begin{equation}\label{eq_frac_wave}
+\nabla( \mathrm{div}\ \mathbf{E})-\nabla^2
\mathbf{E}=\frac{4\pi\mu}{c^2}\frac{\partial \mathbf{j}}{\partial
t}.
\end{equation}

At small times we have ($t\ll 1/\omega_p$)
$$
\frac{\mu\varepsilon_\infty}{c^2}\frac{\partial^2
\mathbf{E}}{\partial
t^2}+\frac{\mu(\varepsilon_s-\varepsilon_\infty)}{c^2}\omega_p^{\alpha\beta}\
_{-\infty}\textsf{D}^{2-\alpha\beta}_t\ \mathbf{E}+
$$
$$
+\nabla( \mathrm{div}\ \mathbf{E})-\nabla^2
\mathbf{E}=\frac{4\pi\mu}{c^2}\frac{\partial \mathbf{j}}{\partial
t}
$$
At large times we have ($t\gg 1/\omega_p$)
$$
\frac{\mu\varepsilon_s}{c^2}\frac{\partial^2 \mathbf{E}}{\partial
t^2}-\frac{\mu(\varepsilon_s-\varepsilon_\infty)}{c^2}\beta\omega_p^{-\alpha}\
_{-\infty}\textsf{D}^{2+\alpha}_t\ \mathbf{E}+
$$
$$
+\nabla( \mathrm{div}\ \mathbf{E})-\nabla^2
\mathbf{E}=\frac{4\pi\mu}{c^2}\frac{\partial \mathbf{j}}{\partial
t}
$$

The wave equation presented above is concordant with equation
obtained by Tarasov (2008-2) from Jonscher's universal law.

\begin{figure}[tb]
\centering
\includegraphics[width=0.41\textwidth]{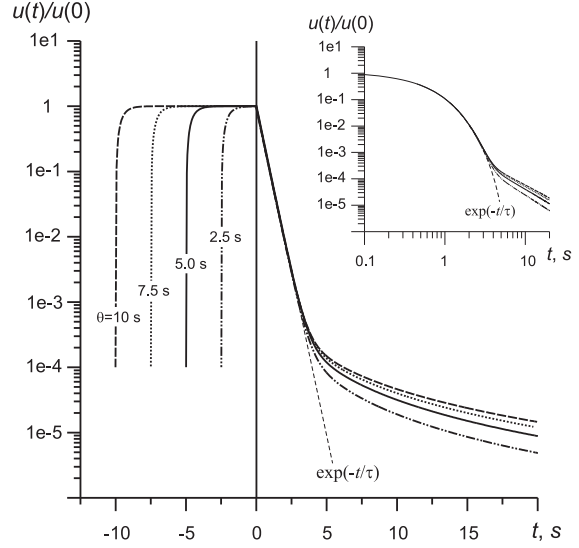}\hspace{1cm}
\caption{\small{Calculated charging-discharging curves for
different $\theta$ ($\beta=1$).}}
\end{figure}

\begin{figure}[tb]
\centering
\includegraphics[width=0.4\textwidth]{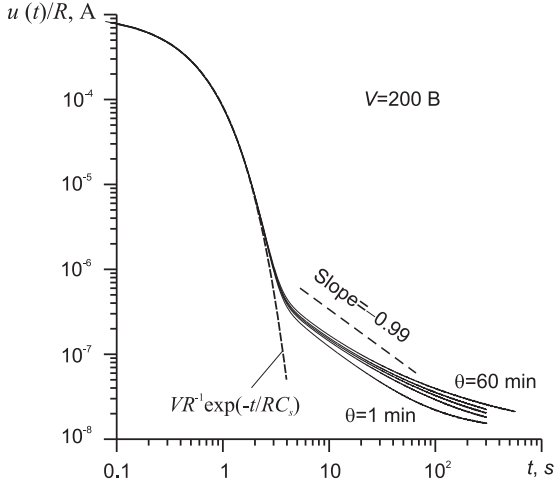}\hspace{1cm}
\caption{\small{Experimental discharging curves for paper
capacitor.}}
\end{figure}

\begin{figure}[tb]
\centering
\includegraphics[width=0.35\textwidth]{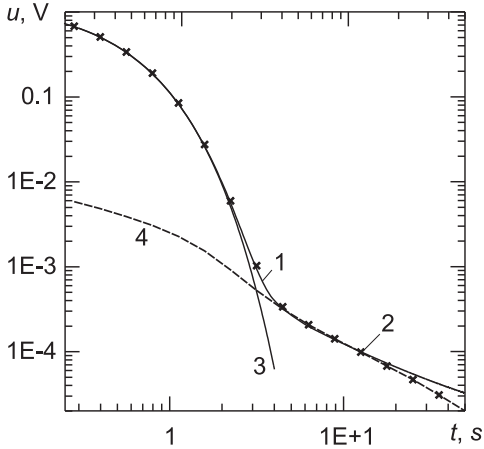}\hspace{1cm}
\caption{\small{Comparison of numerical results (2) with
experimental results (1) for paper capacitor $\theta=300$~ñ,
$\tau\approx 0,4$~c$^{-1}$. Parameter $\alpha$ is equal to 0,998.
3 -- exponential function; 4 -- difference between theoretical
curve and exponential function.}}
\end{figure}

\begin{figure}[tb]
\centering
\includegraphics[width=0.4\textwidth]{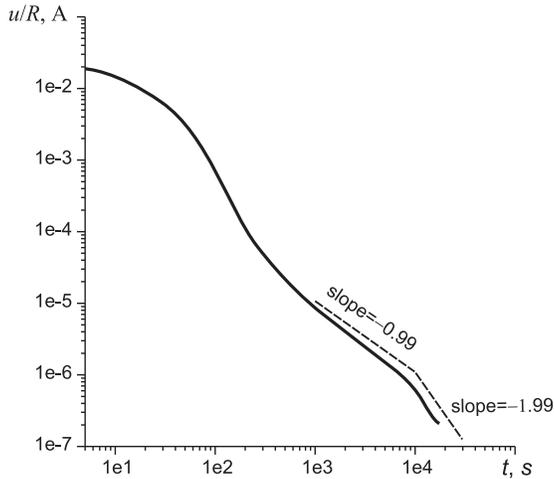}\hspace{1cm}
\caption{{\small Experimental discharging curve for electrolyte
capacitor ($\theta=3600$~ñ).}}
\end{figure}

\section{Comparison of numerical results with experimental data}

We compute charging-discharging curves with the help of
constructed algorithm for $\alpha=0.998$. Charging durations are
$\theta=10$, 7.5, 5, 2.5 seconds. The voltage falls according to
the Debye exponential law for a long time but after some crossover
moment we observe splitting with respect to different values of
$\theta$ and transition to non-Debye power laws. It seems as if
the memory returns to the system after some interval of time. Such
behavior was called in (\cite{Uch:05}) the regeneration memory
effect. When $\alpha=1$ the relaxation follows the Debye law and
does not depend on $\theta$: the memory is absent.

Experimental measurements (see, \cite{Uch:08_3}) were carried out
in the following way. At the first time capacitor was shunted
through the resistor $R$ and ammeter. Necessary bias voltage was
applied by a power source. The capacitor was being charged during
time $\theta$. After charging time $\theta$ capacitor became
shunted again. The capacitor consists of technical paper
impregnated by oil. Its stationary capacitance equals
$2\cdot10^{-6}$~F. For current limitation the resistor
$2\cdot10^5$~Ohm was chosen. Voltage of power source is 200~V.

Thus, experimental data presented in \cite{Uch:08_3} and given in
Fig.~3 confirm the behavior predicted by calculation. Relaxation
does not depend on the charging prehistory in some time domain but
after enough large time process becomes dependent on charging
history.

In Fig. 4 the comparison of calculated discharging process
(points) and experimental data (solid line) is given. It was used
just one parameter adjusted, $\alpha$ ($\beta=1$). Theoretical and
experimental solutions are in a good agreement. Some deviation
occurs at large times.

For charging-discharging process the transition from one power
decaying law ($t^{-\alpha}$) to another ($t^{-\alpha-1}$) is
theoretically predicted in the case of quite large charging times
$\theta$. This transition is quite difficult for observation in
experiment, because it occurs at large times. Nevertheless this
change of exponents in power decaying law has been experimentally
observed for electrolyte capacitor. Charging time was equal to an
hour.

\section{Conclusion}

Let us summarize what we represent in this article. Fractional
relaxation equation (\ref{eq_frac_relax}) and fractional wave
equation (\ref{eq_frac_wave}) for dielectrics with the response
function of the Havriliak-Negami type are considered. The explicit
expression for the fractional operator in these equations is
obtained. The Monte Carlo algorithm for calculation of action of
this operator and of the inverse one is constructed. The algorithm
is derived from the Bochner-Phillips relation for a semigroup
generated by a fractional power of initial infinitesimal operator.
The method is based on averaging procedure over ensembles of
one-sided stable variables.

Relaxation functions calculated numerically according to this
scheme coincide with analytical functions obtained earlier by
other authors. The algorithm represents a numerical way of
calculation in relaxation problems with arbitrary initial and
boundary conditions.

The case, when $\alpha$ is close (but not equal) to 1 and
$\beta=1$, is considered in more details. Numerical calculations
show that in this case does almost not depend on the prehistory
and approximately coincides with  up to some value of argument,
after which the situation changes. Namely, the solutions with
different prehistories reveal different behavior of the inverse
power type on the contrary to the exponential behavior in the
case. Such phenomena named memory regeneration phenomena in
previous publications of the authors are confirmed by experiments
with dielectric capacitors.

\begin{ack}
Authors are grateful to Dr.~S.~Ambrozevich for kindly presented
experimental results and to Prof. V. Uchaikin for fruitful
discussions.
\end{ack}

\end{document}